\documentclass[fleqn,usenatbib]{mnras}

\usepackage{newtxtext,newtxmath}

\usepackage[T1]{fontenc}

\DeclareRobustCommand{\VAN}[3]{#2}
\let\VANthebibliography\thebibliography
\def\thebibliography{\DeclareRobustCommand{\VAN}[3]{##3}\VANthebibliography}

\usepackage{graphicx}	
\usepackage{amsmath}	

\newcommand{\rmicron}{$\, \rm \mu m$}

\title[Hot Dust at High Redshift]{The early Universe was dust-rich and extremely hot}

\author[Marco P. Viero et al.]{
Marco P. Viero,$^{1}$\thanks{E-mail: marco.viero@caltech.edu}
Guochao Sun,$^{1}$
Dongwoo T. Chung,$^{2,3}$
Lorenzo Moncelsi,$^{1}$
Sam S. Condon$^{1}$
\\
$^{1}$California Institute of Technology, 
1200 East California Boulevard, 
Pasadena, CA, 91125, USA\\
$^{2}$Canadian Institute for Theoretical Astrophysics, 
University of Toronto,  
60 St. George Street, 
Toronto, ON, M5S 3H8, Canada\\
$^{3}$Dunlap Institute for Astronomy and Astrophysics, 
University of Toronto, 
50 St. George Street, 
Toronto, ON, M5S 3H4, Canada
}

\date{Accepted XXX. Received YYY; in original form ZZZ}

\pubyear{2022}

\begin{document}
\label{firstpage}
\pagerange{\pageref{firstpage}--\pageref{lastpage}}
\maketitle

\begin{abstract}
We investigate the dust properties and star-formation signature of galaxies in the early Universe 
by stacking 111\,227 objects in the recently released COSMOS catalogue on maps at wavelengths bracketing the peak of warmed dust emission.  
We find an elevated far-infrared luminosity density to redshift 8, indicating abundant dust in the early Universe.    
We further find an increase of dust temperature with redshift, reaching $100\pm 12\ \rm K$ at  $z\sim 7$, suggesting either the presence of silicate rich dust originating from Population II stars, or sources of heating beyond simply young hot stars.  
Lastly, we try to understand how these objects have been missed in previous surveys, and how to design observations to target them.  
All code, links to the data, and instructions to reproduce this research in full are located at \url{https://github.com/marcoviero/simstack3/}.
\end{abstract}




\section{Introduction}

The history of star formation at high redshift has 
always had a dust problem. Star formation is well traced by ultraviolet emission, but a large fraction of ultraviolet light is obscured by dust, thus proving it to be an incomplete estimator.  But dust also {\it emits} light, re-radiating the ultraviolet energy it absorbed as roughly a blackbody at far-infrared (FIR) wavelengths.  
To measure the blackbody is to measure star formation; to measure it across cosmic time is to characterise cosmic star-formation history, and maybe learn something about the evolution of the different galaxy populations.

Fundamental questions about the early Universe still remain, such as: 
was it dust rich or dust poor?  How much does obscured star formation contribute to the star-formation rate density?  
Do evolutionary trends of galaxies with redshift seen locally (e.g. temperature, specific star-formation rate) continue indefinitely, or do they plateau, or even turn over?  
Just how active are AGNs in this period?  And how much ionizing radiation is escaping the dark curtain of dust {\it exactly}?

The catch is that measuring the far-infrared spectrum of a galaxy at high redshift is challenging.  A modified blackbody spectrum (or spectral-energy distribution; SED) with a temperature of 18\,K peaks in the infrared at a rest-frame of roughly 160\rmicron, or between 400--600\rmicron{} in the observer frame if emitted at the peak of star formation, $z\sim2$--3.  To fully bracket that peak requires combining observations from space telescopes like {\it Spitzer} and {\it Herschel}, 
and ground based observatories like the JCMT, LMT, and ALMA.  
FIR space telescopes are wonderful except for one drawback: source confusion.  The primary beams are large enough to contain tens (or hundreds) of galaxies, complicating optical counterpart identification.  Ground-based FIR telescopes are wonderful except for a different drawback: the atmosphere, which is to infrared light what dust is to optical light.  

Until we are able to put an actively-cooled, $50\, \rm m$ mirror in space, the solution is stacking.  
Galaxies with similar physical properties estimated from optical photometry (e.g. redshift, stellar-mass, star-forming activity) have, on average, similar infrared properties \citep[e.g. infrared luminosity, dust temperature;][]{schreiber2018}. 
\citet{viero2013} developed an algorithm to stack galaxies, grouped by their similarity, simultaneously \citep[see also][]{kurczynski2010}, thus providing a way to overcome the bias inherent to stacking in highly confused images.

The techniques in this paper are not new. 
The reason we are only now able to address these questions is because of newly available catalogues released to the public from COSMOS \citep{scoville2007}.  
We combine the COSMOS2020 catalogue \citep{weaver2022} with infrared/submillimetre maps publicly available through the Herschel Extragalactic Legacy Project (HELP) and SCUBA-2 Cosmology Legacy Survey (S2CLS), and a newly released, {\sc python3}-compatible version of {\sc simstack}, to measure the far-infrared SEDs from redshift 0 to 10 and stellar-masses of $\log(M/\mathrm{M_{\odot}})=9.5$ to 12.

We find that the early Universe was dust-rich, and unusually hot. 
Effective dust temperatures increase with redshift, following established relationships to $z=4$, but then rise more rapidly to $z\sim 10$.  
Similarly, we find that the far-infrared luminosity density ($\rho_\mathrm{LIR}$) tracks the established star-formation rate density (SFRD or $\rho_\mathrm{SFR}$) to $z=2$, but then decouples from it, remaining flat to $z=8$. 
Given the unusual nature of the hot dust at $z > 4$, we explore the mechanisms that could be responsible for this heating, how this could have been missed previously, and strategies for follow-up observations.

We assume a \citet{chabrier2003} initial-mass function (IMF) and Planck18 cosmology 
\citep{2020A&A...641A...6P}, 
with $\Omega_{\rm M}=0.315$,  $\Omega_{\Lambda}=0.685$, $H_0=67.4\, \rm km\,s^{-1}\, Mpc^{-1}$, $\sigma_8=0.811$, and $\tau = 0.054$.
All code and instructions to reproduce these results can be found at \url{https://github.com/marcoviero/simstack3/} and 
at the DOI: \url{https://doi.org/10.5281/zenodo.6792395}.

\section{Data} \label{sec:data}

\subsection{COSMOS2020 Catalogue} \label{sec:catalog}

Our analysis is performed on data in the COSMOS\footnote{\url{https://cosmos.astro.caltech.edu}} field \citep{scoville2007}, centred at $\rm 10^h00^m26^s, 2^{\circ}13^{\prime}00^{\prime\prime}$, and covering 1.606\, deg$^2$ after masking. We use the recently released COSMOS2020 catalogue \citep{weaver2022}, whose $K_s = 25.2\, \rm (AB)$-selected sample is an improvement of nearly two magnitudes over previous releases used for stacking \citep{muzzin2013}.  
This extra depth enables access to objects reaching redshift of 10, and completeness improvements of $\sim 2\, \rm dex$ in stellar mass.  

The COSMOS2020 release comes with two options for photometry; we use {\sc FARMER}/{\sc LePhare}, generally understood to be superior for faint sources at high redshift.  
Following \citet{weaver2022}, we split the population into star-forming and quiescent using the $NUV-r$ versus $R-J$ selection \citep{ilbert2013}.

\subsection{Maps} \label{sec:maps}

Maps are selected to bracket the rest-frame dust SED at $z=0$--10. Unless otherwise noted, they were obtained from the Herschel Extragalactic Legacy Project (HELP\footnote{\url{https://herschel.sussex.ac.uk}}). 

The {\it Spitzer}/MIPS map at 24\rmicron\ is a Spitzer Enhanced Imaging Product (SEIP). 
It has an rms of $27.9\, \rm \mu Jy$, point-spread function (PSF) full-width at half maximum (FWHM) of 6.32\,arcsec, and aperture/colour corrections equaling 1.24 \citep{bethermin2010}. Maps are converted from MJy\,sr$^{-1}$ to Jy\,beam$^{-1}$ by dividing the map by the beam solid angle, $1.55\times10^{-9}\, \rm sr$.  

\emph{Herschel}/PACS maps at 100 and 160\rmicron\ are from the PACS Evolutionary Probe \citep[PEP;][]{poglitsch2010}.  
They have rms values of $0.10$ and $2.11\, \rm mJy$, and FWHM of 7.5 and 11.3\,arcsec. PACS maps are also converted from MJy\,sr$^{-1}$ to Jy\,beam$^{-1}$, with beam solid angles of $2.03\times10^{-9}$ and  $4.66\times10^{-9}\, \rm sr$. Calibration uncertainty is 4\%.  

\emph{Herschel}/SPIRE maps at 250, 350, and $500\, \rm \mu m$ are from the Herschel Multi-tiered Extragalactic Survey \citep[HerMES;][]{oliver2012}, with instrumental depths of 15.9, 13.3, and 19.1\,mJy ($5\sigma$); 
FWHM are 17.6, 23.9, and 35.2\,arcsec.  Calibration uncertainty is 5\%. 

The $850\,\rm \mu m$ map is a product of 
S2CLS\footnote{\url{https://zenodo.org/record/57792}} \citep{geach2017}.
Map depth is 1.6\,mJy\, beam$^{-1}$ ($1\, \sigma$), and FWHM 12.1\,arcsec. 
Note, we specifically select the non-match-filtered (NMF) maps, which retain the unresolved fluctuations on which {\sc simstack} relies.

\section{Method} \label{sec:method}

\subsection{Unbiased Stacking with {\sc simstack}} \label{sec:simstack}

Our stacking analysis is based on {\sc simstack}, an algorithm designed to overcome the biases inherent with stacking on images dominated by confusion noise.  
Briefly, {\sc simstack} uses the 3D positions in a catalogue to construct and fit a model to a map. 
The model is made up of layers representing bins that have been defined by the user (e.g. by stellar-mass, redshift, etc.), and whose best-fittings are interpreted as the average flux density of the binned objects.  
The software for stacking used here is the same one detailed in \citet{viero2013} with some minor modifications.  

The first difference, following \citet{sun2018}, is that instead of stacking in redshift slices, now all redshift bins are also stacked simultaneously.  
This change is to minimize the potential bias that may arise from objects outside the bin boundaries, or from objects correlated with interloping sources.

The next difference is motivated by \citet{duivenvoorden2020}, who showed that stacked flux densities can be biased by holes in the catalogue coverage caused by extremely bright objects like stars. 
Their solution is the addition of a flat foreground layer, which is now an optional flag (default=True) in {\sc simstack}. 

The last change concerns the bootstrap estimation.  Previously, the bootstrap consisted of randomly sampling the catalogue (with replacement), which is robust in most scenarios, but in this case actually biases the fluxes low.  
The modification satisfies the central principle of {\sc simstack}, that all correlated objects must be stacked at the same time to prevent biasing the estimate, by randomly splitting each bin in two, with a 80:20 ratio, and stacking the entire set simultaneously.  
The 80\% bins are retained as the bootstrapped flux, and the 20\% bins are examined for outliers.  
The drawback with this method is that the memory required to perform the stack is doubled, requiring either a more powerful computer, or splitting the bootstrap into chunks.

\begin{figure*}
\includegraphics[width=1.0\textwidth]{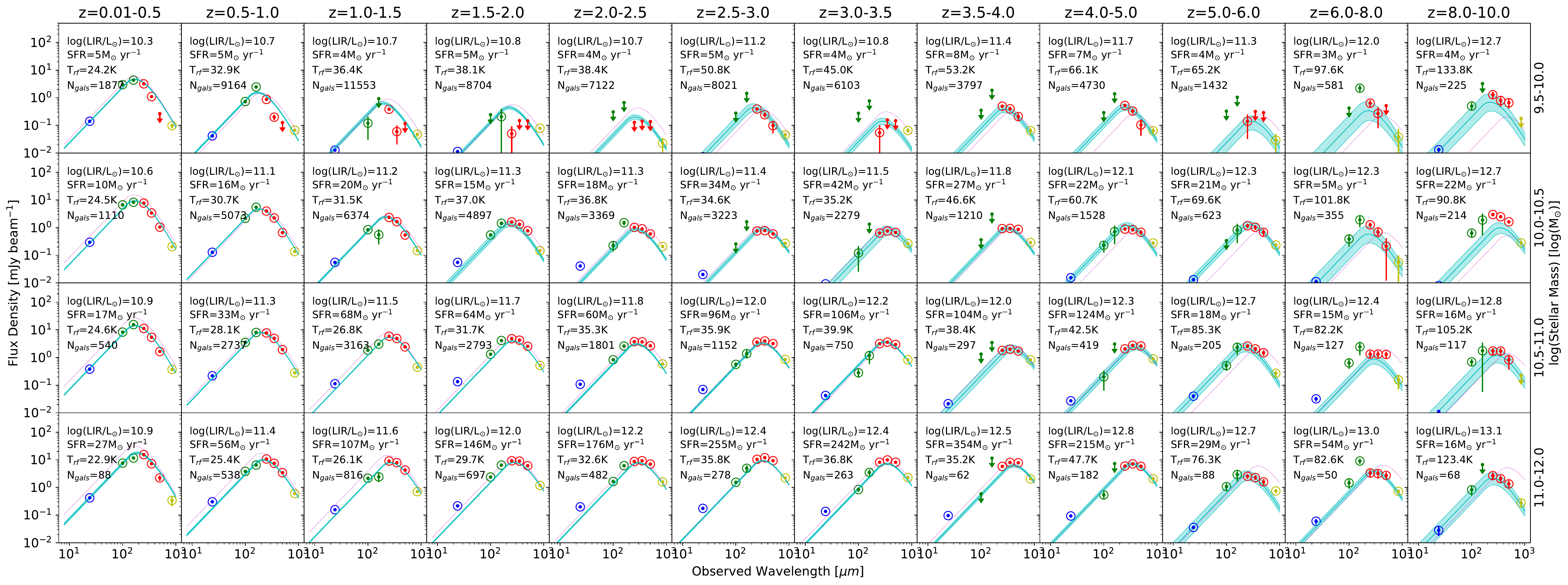}
\caption{Best-fitting SEDs of star-forming galaxies, binned by redshift (increasing left to right) and stellar mass (increasing top to bottom).  
Data at 24\rmicron\ (blue circles) are from \emph{Spitzer}/MIPS; at 100 and 160\rmicron\ (green circles) from \emph{Herschel}/PACS; at 250, 350, and 500\rmicron\ (red circles) from \emph{Herschel}/SPIRE; and at 850\rmicron\ (yellow circles) from SCUBA2.  
Error bars represent $1\, \sigma$ Gaussian uncertainties, derived with the bootstrap estimator, on 150 iterations.  Non-detections are shown at $3\, \sigma$ upper limits.
The SED model is a hybrid blackbody with fixed emissivity index $\beta=1.8$ on the Rayleigh-Jeans side, and fixed power-law approximation slope $\alpha = 2$ on the Wien side.   
MCMC fits (cyan) show the 50th percentile line bracketed by a shaded region marking the 25th and 75th percentiles. 
For comparison we also show hypothetical main-sequence galaxy SEDs (thin magenta lines), extrapolated from the \citet{schreiber2015} model at $z < 4$.  
\label{fig:mcmcs}}
\end{figure*}

\subsection{Spectral Energy Distribution Fitting} \label{sec:mcmc}

To obtain SEDs from stacked flux densities we fit them with a hybrid blackbody, substituting the mid-infrared Wien side with a power law, $\alpha$, with fixed slope of 2.0, and emissivity index $\beta$ fixed at 1.8 \citep[e.g.][]{casey2012}. 
Fits then have two free parameters, amplitude and observed temperature.  We sample the parameter space with {\sc emcee} for {\sc python} \citep{foreman2013}, a user-friendly, affine-invariant ensemble sampler for Markov chain Monte Carlo (MCMC).  
We adopt a flat prior with generous upper and lower limits for amplitude and temperature. For five bins at $z=1.5$--2 and $z=2.0$--2.5, we add a prior on temperature to prevent the solution from preferring local minima at higher temperature. 
Additionally, we account for PAH emission lines contaminating the 24\rmicron\ flux density by artificially inflating the MIPS errors over the redshift range  $0.5<z<3$. 
Full covariances are derived from bootstraps to account for the highly correlated nature of bands near to one another in wavelength.  
Flux density measurements consistent with zero are treated as upper limits, whose contribution to the log-likelihood is via a survival function \citep{sawicki2012}.  
 We use 15\,000 samples, discarding the first 3000.  

Infrared luminosities (LIR) are estimated as the integral of $\nu I_{\nu}$ from rest-frame 8 to 1000\rmicron\ \citep{casey2012}, and na\"{i}vely converted to the star-formation rate (SFR) using the \citet{kennicutt1998} relation, with a 0.23\,dex correction to convert to a \citet{chabrier2003} IMF. 
Error bars account for additional uncertainties in the luminosity distances arising from the photometric redshifts.   

We also estimate the SFR via the rest-frame 850\rmicron\ luminosity $L_{850}$. As a proxy for the interstellar medium gas mass, it is insensitive to dust temperature $T_\mathrm{d}$ with a relatively constant luminosity--mass ratio, and is also a reliable SFR estimator \citep{scoville2016}.  

\subsection{Null Tests and Other Consistency Checks} \label{sec:null}

To check that the signal we measure is not noise, we redo the stack but randomly shuffle source positions. 
We find that the measured flux densities are centred around zero, with uncertainties consistent with the rms in the maps. 
We further check that the measured signal is not dominated by a few bright sources by inspecting the distribution of the 20\% group of the bootstrap split, finding their relative variance is proportional to their relative size. 

Next, we carefully examine the impact that low-$z$ interlopers have on the stacked SED shape at $z > 4$ using simulated catalogs/maps generated with the {\sc pysides} 
 simulation \citep{bethermin2010}.
We use the redshift probability distributions $P(z)$ provided by the COSMOS team 
to i) draw a redshift for each object; ii) look for the closest counterpart in the {\sc sides} catalog; iii) bin and average their flux densities; iv) and compare the best-fit SEDs.  We repeat this 100 times (notebooks and downloads of the simulation results can be found in the same GitHub repository as for the main results).

Objects at higher redshift or with lower stellar masses have broader $P(z)$, sometimes even double-peaked; provided they are reliable probabilistic estimators of the true redshifts, this method should give a reasonable representation of the outlier bias on the mean SED.  
We find that the overall bias in rest-frame dust temperature is small, but that the scatter is large, particularly at $z > 5$.  We apply these excess uncertainties to the errors in the dust temperatures.

Finally, we check that the cumulative flux density is consistent with the cosmic infrared background \citep[e.g.][]{dole2006}, and that the contributions from different redshift and stellar-mass bins agree with \citet{viero2013}.

\section{Results} \label{sec:results}

The full set of SEDs for star-forming galaxies are shown in Figure~\ref{fig:mcmcs}.
The best fits are in blue, with blue shaded regions outlining the 25th and 75th quartiles of the full chain of MCMC samples.  
Overlaid in magenta are SEDs drawn from the star-forming main sequence of \citet{schreiber2015}, as described in \S~\ref{sec:null}, showing a clear departure from existing trends at $z=0$--4.

\begin{figure}
\includegraphics[width=\columnwidth]{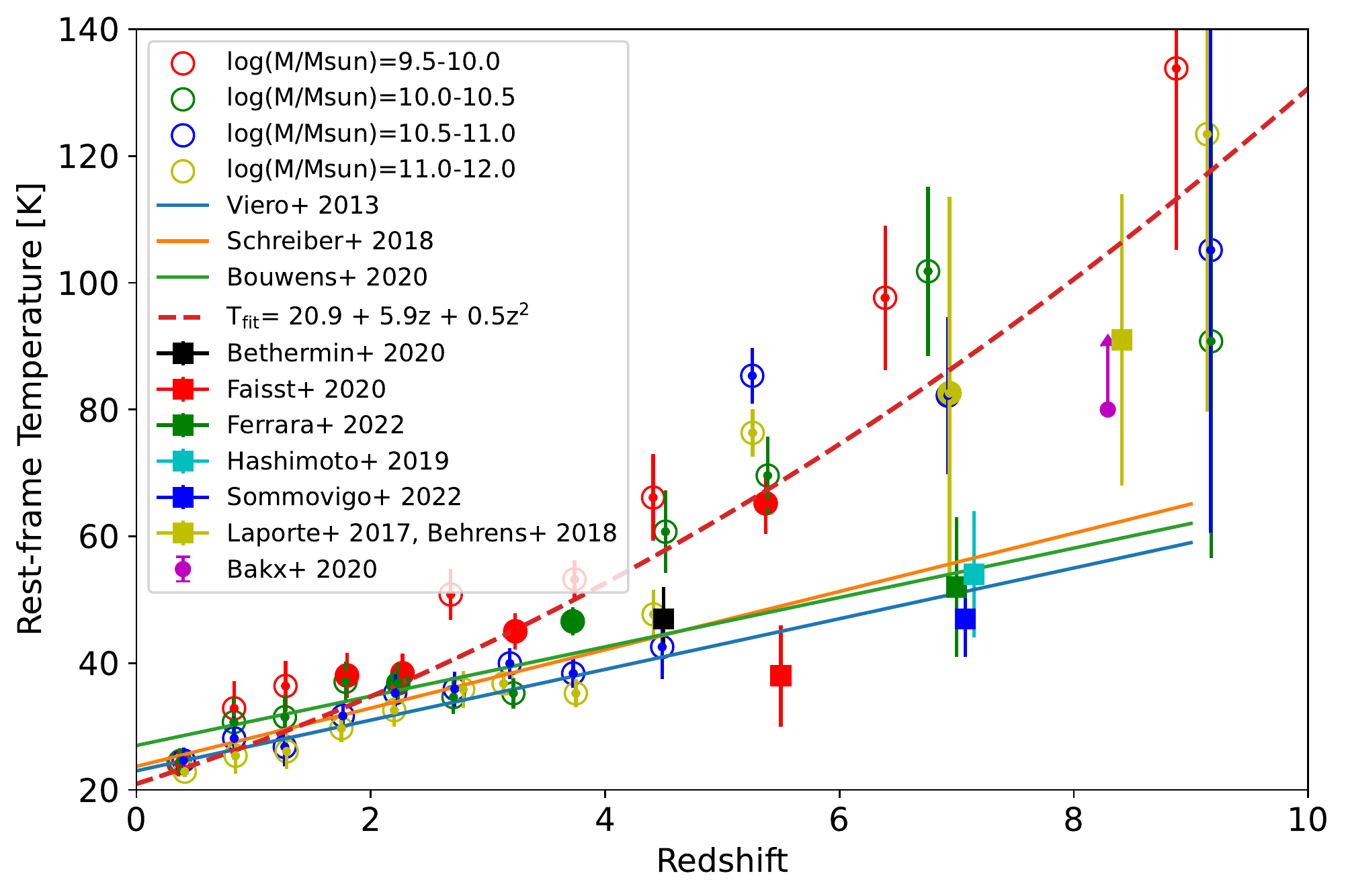}
\caption{Rest-frame dust temperatures from best-fittings to SEDs plotted as open circles. A polynomial fit for $T(z)$ is shown as a dashed red line. Closed circles are the five bins fit with priors, and are not included in the fit. Positions on the redshift axis correspond to median redshifts of the objects in the bins.  Also plotted are $T_\mathrm{d}$--$z$ relationships \citep{viero2013,schreiber2018,bouwens2020}, extrapolated to $z=9$. Temperatures of individual objects include \citet{hashimoto2019}, \citet{faisst2020}, \citet{sommovigo2022}, \citet{bethermin2020}, and \citet{laporte2017}/\citet{behrens2018}, and lower limit from \citet{bakx2020}.  \label{fig:tdust} }
\end{figure}

\subsection{Dust Temperature} \label{sec:tdust}

We find dust temperatures increasing with redshift (Figure~\ref{fig:tdust}), in agreement with previous relationships found to redshift 4 \citep{viero2013, schreiber2018, bouwens2020}.  
At $z > 4$, we find temperatures diverge from those relationships. 
Temperatures are incompatible with other estimates that follow (extrapolated) existing relationships \citep[e.g.][]{bethermin2020,faisst2020, ferrara2022, hashimoto2019,sommovigo2022}.  
They are in better agreement with the lower limit of \citet{bakx2020}, and estimate of \citet{behrens2018} based on data from~\citet{laporte2017}; these findings had been considered outliers. 
We model the $T_\mathrm{d}$--$z$ relationship with a polynomial: 
\begin{equation}
    T_\mathrm{d}(z) = (20.9\pm3.5) + (5.9\pm 1.9)\, z + (0.5\pm 0.2)\, z^2.
\end{equation}

Na\"{i}ve explanations for these findings \citep[after considering heating by the cosmic microwave background;][]{dacunha2013} are inadequate. Ambient heating by the elevated cosmic microwave background is subdominant in these objects \citep{dacunha2013}.  
Further, at $z\sim 7$, assuming typical, steady-state dust grains and an interstellar radiation field with total energy density $U\gtrsim1500\,\rm{eV\,cm^{-3}}$, to supply enough energy by star formation would require an SFR of greater than $1000\, \rm M_{\odot}\,\mathrm{yr^{-1}}$. Even for a halo as massive (and rare) as $\rm 10^{13}\,M_{\odot}$, this entails an high, order-of-unity star formation efficiency \citep{sun2016}. 

However, the dust composition of galaxies at these redshifts is very different from local \citep[e.g.][]{behrens2018}. Multiple dust components (e.g. cold vs.\@ warm) may be present \citep{strandet2017,liang2019}. As pointed out by \citet{derossi2018}, dust in primeval galaxies (Pop II) is typically younger than $400\, \rm Myr$; silicate rich, being produced by low metallicity AGB stars \citep{ventura2012}, with poor emission efficiency at low and high wavelengths \citep{koike2003}; and located in environments near to the heating source \citep{spilker2016}.
AGNs are another likely significant source of heating \citep{lyu2016}, whose host galaxies are very compact \citep{decarli2018}, with SFRs of $1000-5000\, \rm M_{\odot}\, yr^{-1}$ \citep[and SFR surface densities reaching $300-2000\, \rm M_{\odot} \, yr^{-1} \, kpc^{-2}$;][]{lyu2016}. 
A fuller treatment must specifically consider the nature of galaxies in the early Universe.

\begin{figure}
\includegraphics[width=\columnwidth]{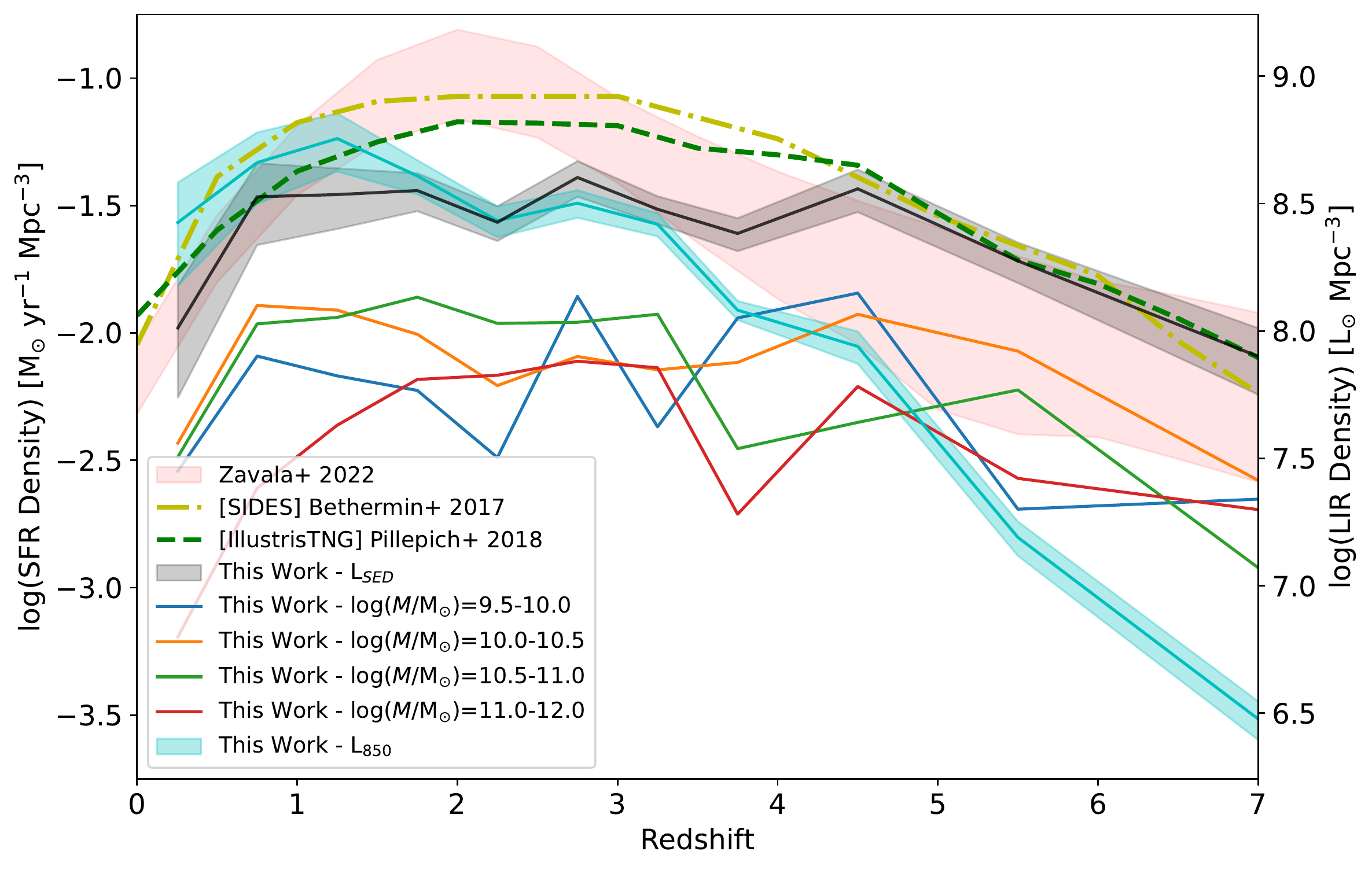}
\caption{Star-formation rate density vs.\@ redshift, binned by stellar mass (coloured lines), and summed (black line with 1\,$\sigma$ shaded region). 
Star-formation rates are converted from far-infrared luminosities -- derived from the integral of intensities from (rest-frame) 8 to 1000\rmicron -- by scaling with the \citet{kennicutt1998} relation.  
We find we are in broad agreement with a selection of the latest $\rm SFRD$ from models \citep{bethermin2017, pillepich2018} and measurements \citep{zavala2021}. 
Also plotted is the rest-frame 850\rmicron-derived star-formation rate density as a light blue line and shaded region ($1\, \sigma$). 
The LIR-derived SFRD tracks 850\rmicron-derived relation to $z=4$, but then they diverge, similar to the dust temperature behaviour, suggesting that the FIR luminosity may not be a good tracer of star-formation at high redshift.  
\label{fig:sfrd}}
\end{figure}

\subsection{The LIR and Star-Formation Rate Density} \label{sec:sfrd}

We estimate the far-infrared luminosity density ($\rho_\mathrm{LIR}$) as a number-weighted sum of the infrared luminosities in different stellar-mass bins derived above, 
\begin{equation}
	\rho_\mathrm{LIR}  =\frac{1}{V} \displaystyle \sum^m _{i=0} L_i N_i, ~~
	\delta \rho_\mathrm{LIR}^2  = \frac{1}{V^2} \displaystyle \sum^m _{i=0} (N_i \delta L_i)^2 + (L_i \delta N_i)^2, 
\end{equation}
where $\delta N^2_i = N^2\epsilon^2_{CV}$, and $\epsilon_{CV}$ is the cosmic variance term \citep{moster2011}. 
Similarly, we estimate the SFR density ($\rho_\mathrm{SFR}$) from SFRs derived from rest-frame 850\rmicron\ flux densities. 

In Figure~\ref{fig:sfrd} we show the $\rho_\mathrm{LIR}$ as a grey line and shaded region, which when converted to $\rho_\mathrm{SFR}$ is in good agreement with previous measurements \citep[e.g.][]{bouwens2020, zavala2021} and models \citep[][]{bethermin2017, pillepich2018}  to $z=4$, but then remains elevated, while the models decline.    
Given our finding of hot dust at $z>4$ this excess is not unexpected, as we know that hot dust SEDs are poor estimators of star-formation rates \citep{bouwens2016}.   
On the other hand, the 850\rmicron-derived $\rho_{SFR}$, shown as a blue line and shaded region, exhibits the expected decline after peaking at $z=1-2$. 
Note, we exclude the $z=8-10$ bins, whose galaxy abundances suggest a large contribution from interlopers.  

It is reasonable to ask how so much of the $\rm SFRD$ can be recovered when the incompleteness levels of the catalogue at $z>4$ must be high.  
The reason is because we are stacking on images with large PSFs, which would result in a bias from galaxies that were not in, but were correlated with, those in the catalogue. 

The high-$z$ SFRD has significant consequences for the evolution of the early Universe \citep{robertson2021}; leveraging statistical techniques like stacking to access the full unresolved population \citep[see][]{viero2015,cheng2021} will complement traditional methods to provide a complete picture of star-formation activity.

\section{Discussion} \label{sec:discussion}
How could so much hot dust in the early Universe have been missed?  In hindsight, all the clues were there:  
there was a dearth of high-redshift dusty galaxies \citep{casey2018} that hot dust could explain \citep{sommovigo2020}. 
There was known tension in low IRX values observed for given $\beta$ at high redshift \citep{capak2015} that could be alleviated with warmer dust \citep{bouwens2016}. 
There were unusually high dust-mass estimates that challenged dust production time-scales \citep{lesniewska2019}, which is resolved with LIR coming from hotter dust \citep{bakx2020}.  
And then there were simple models showing that ALMA observations tended to miss galaxies with high-temperature dust \citep{chen2022}.  

So why has the idea of \emph{hot} dust at high redshift not gained traction?
First, a handful of hot dusty galaxies at $z> 6$ \emph{had} been observed \citep[e.g.][]{behrens2018, bakx2020}, but many others were found to be consistent with extrapolations to previous relationships \citep[e.g.][]{hashimoto2019, faisst2017, sommovigo2020, ferrara2022}, so that the behaviour of the $T_{\rm d}-z$ relationship -- increasing or plateauing -- was still unclear \citep{faisst2017, sommovigo2020}. 
However, those temperature estimates were derived from measurements with ALMA that did not actually bracket the peak of the SED (largely staying below Band 8, so no shorter than 600\rmicron{}). If in fact these objects were much hotter, no adequate prior to extrapolate model SEDs even existed. 

Alongside hot dust is an elevated infrared-luminosity density at $z>4$, which appears to decouple from the cold-dust derived SFRD.  
The source of this excess heating is likely some combination of young, Pop II stars and silicate rich dust, and AGN, but in what combination, and is that enough? 
Such details impact our understanding of reionization time-scales, or of the presence of quiescent galaxies at high redshift \citep{glazebrook2017}. 

Understanding the nature of the dust, and the extreme source of heating -- is it stars, AGNs, PBHs? -- ranks high on the list of fundamental questions going forward. 

\section{Conclusion} \label{sec:conclusion}
We estimate far-infrared luminosities and dust temperatures from galaxies at $z=0$--10, by stacking 111\,227 objects from the COSMOS2020/{\sc Farmer} catalogue on maps in the far-infrared/submillimetre.
We find agreement with previous measurements to $z=4$, but beyond this find elevated temperatures and  luminosities, indicating an early Universe filled with hot dust.  
The temperatures measured suggest a direct detection of Pop II galaxies -- compact, close-packed, and silicate rich -- whose SEDs carry the signature of localized hot dust.   

If confirmed, this will have far-reaching consequences.  It will inform future probes into the epoch of reionization, whether directly with the \emph{James Webb Space Telescope (JWST)}, or via statistics of spectral lines like CO and [\ion{C}{ii}] with line-intensity mapping experiments like COMAP~\citep{clearly22}, TIME~\citep{Sun21}, or CONCERTO~\citep{CONCERTO}.  In particular, a warm dust background could present problems for  [\ion{C}{ii}] experiments by decreasing the line-continuum contrast.

Clearly, follow-up observations targeting the peak of hot dust emission are in order.  
ALMA band 10 will probe nearer to the peak than band 9 observations; targeting the most massive and reliable subset of the $z>7$ sample should achieve detections with reasonable signal-to-noise, but still will not fully bracket the peak. 

Long term, dedicated observatories targeting the peak of warm dust SEDs in the submillimetre will be necessary to complement deep optical/NIR observations. On the horizon are up-to-50 m class ground-based single-dish observatories like CCAT-prime/FYST~\citep{FYST}, AtLAST~\citep{klassen2020}, and the Japanese 30 m Antarctic THz telescope intended for New Dome Fuji~\citep{chen2022}, as well as space missions \citep[e.g. \emph{Origins Space Telescope};][]{wiedner2020}.  


\section*{Acknowledgements}

The authors are grateful to the referee for their careful reading and constructive suggestions, thank you.
DTC holds a CITA/Dunlap Institute postdoctoral fellowship.
The Dunlap Institute is funded through an endowment established by the David Dunlap family and the University of Toronto. 
The University of Toronto operates on the traditional land of the Huron-Wendat, the Seneca, and most recently, the Mississaugas of the Credit Rive. 

\section*{Data Availability}

The {\sc simstack} package needed to reproduce these results, and Jupyter Notebooks guiding the user through each step, are available  at \url{https://github.com/marcoviero/simstack3/tree/main/viero2022/} and 
at the DOI: \url{https://doi.org/10.5281/zenodo.6792395}.  



\bibliographystyle{mnras}
\bibliography{v2022} 



\bsp	
\label{lastpage}
\end{document}